\documentclass[sigconf, nonacm]{acmart}

\usepackage{listings}
\usepackage{pifont}

\usepackage{tikz}
\usetikzlibrary{
  arrows,                        
  backgrounds,                        
  ext.paths.ortho,                    
  ext.positioning-plus,               
  ext.node-families.shapes.geometric, 
  shapes.geometric,                   
  calc}      

\tikzstyle{standard} = [rectangle, draw, rounded corners, text width=3.7cm, minimum height = 3.6cm]
\tikzstyle{wide} = [rectangle, draw, rounded corners, text width=8.31cm, minimum height = 3.0cm]
\tikzstyle{connector} = [draw, -latex', line width=2pt]
\tikzset{
    arrow/.style={thick, ->, >=stealth},
  header node/.style={
    font =\strut\small,
    text depth=+.3ex, fill=white, draw},
  header/.style={%
    inner ysep=+1.5em,
    append after command={
      \pgfextra{\let\TikZlastnode\tikzlastnode}
      node [header node] (header-\TikZlastnode) at (\TikZlastnode.north) {#1}
      node [fit=(\TikZlastnode)(header-\TikZlastnode), inner sep=0pt]
           (h-\TikZlastnode) {}
    }
  },
}

\begin{document}

\title{Assessing Crime Disclosure Patterns in a Large-Scale Cybercrime Forum}

\author{Raphael Hoheisel}
\orcid{0000-0002-3621-425X}
\affiliation{%
  \institution{University of Twente}
  \country{Netherlands}
}

\author{Tom Meurs}
\orcid{0000-0003-0963-5232}
\affiliation{%
  \institution{Dutch Police}
  \country{Netherlands}
}

\author{Jai Wientjes}
\orcid{0009-0005-7064-907X}
\affiliation{%
  \institution{Dutch Police}
  \country{Netherlands}
}

\author{Marianne Junger}
\orcid{0000-0002-9515-9860}
\affiliation{%
 \institution{University of Twente}
  \country{Netherlands}}

\author{Abhishta Abhishta}
\orcid{0000-0001-7122-3103}
\affiliation{%
  \institution{University of Twente}
  \country{Netherlands}}

\author{Masarah Paquet-Clouston}
\orcid{0009-0005-4934-0019}
\affiliation{%
  \institution{Université de Montréal}
  \city{Montréal}
  \institution{Complexity Science Hub}
  \country{Austria}}

\renewcommand{\shortauthors}{Hoheisel et al.}

\begin{abstract}
    Cybercrime forums play a central role in the cybercrime ecosystem, serving as hubs for the exchange of illicit goods, services, and knowledge. Previous studies have explored the market and social structures of these forums, but less is known about the behavioral dynamics of users, particularly regarding participants’ disclosure of criminal activity. This study provides the first large-scale assessment of crime disclosure patterns in a major cybercrime forum, analysing over 3.5 million posts from nearly 300k users. Using a three-level classification scheme (benign, grey, and crime) and a scalable labelling pipeline powered by large language models (LLMs), we measure the level of crime disclosure present in initial posts, analyse how participants switch between levels, and assess how crime disclosure behavior relates to private communications. Our results show that crime disclosure is relatively normative: one quarter of initial posts include explicit crime-related content, and more than one third of users disclose criminal activity at least once in their initial posts. At the same time, most participants show restraint, with over two-thirds posting only benign or grey content and typically escalating disclosure gradually. Grey initial posts are particularly prominent, indicating that many users avoid overt statements and instead anchor their activity in ambiguous content. The study highlights the value of LLM-based text classification and Markov chain modelling for capturing crime disclosure patterns, offering insights for law enforcement efforts aimed at distinguishing benign, grey, and criminal content in cybercrime forums.
\end{abstract}




\maketitle

\section{Introduction}

    Cybercrime forums play a pivotal role in the online criminal ecosystem. They serve as hubs where users exchange information, establish connections and engage in the trade of illicit goods and services~\cite{leukfeldt_use_2017}. Given their importance in facilitating cybercrime, a substantial amount of studies examining these forums exists. Their focus ranges across several topics, including, to name a few, understanding how trust is built ~\cite{holt_examining_2013, dupont_darkode_2017}, analysing how the forums' markets function~\cite{yip_why_2013, moore_economics_2009, boekhout_early_2024}, and how they support the growth of criminal networks and recruitment \cite{leukfeldt_use_2017}.  
    
    Nevertheless, although these forums serve as hubs for cybercrime, discussions often extend far beyond criminal topics, encompassing topics such as music, mental health, and other benign interests~\cite{talas_hackers_2023, man_autism_2023, dupont_darkode_2017}.  Moreover, when examining user activity, prior studies have shown that a small group of highly engaged users typically drives much of the forum’s discussions, while the vast majority of users show limited participation~\cite{pastrana_characterizing_2018, paquet-clouston_assessing_2018}. 
    
    Given that many users contribute minimally on these forums and discussions encompass a wide range of topics, it remains unclear how much crime-related activity actually occurs on cybercrime forums. This is further complicated by the fact that, due to the  commoditization and specialization features of the online criminal ecosystem~\cite{moore_economics_2009, wegberg_plug_2018, collier_cybercrime_2020, collier_cybercrime_2021}, many cybercrime tasks resemble legitimate IT work, such as infrastructure maintenance~\cite{bijlenga_criminals_2018}. This creates a ``grey area'' where discussions are not directly criminal, but occupy an ambiguous space in between, making it difficult to distinguish benign participation from actual criminal engagement~\cite{paquet-clouston_extending_2025}. Nevertheless, forum participation does not inherently imply criminal engagement, and legitimate doubt often remains, an issue often overlooked by previous researchers, that warrants further investigation. 
    
    Alternatively, one can view cybercrime forum participation as akin to being in a state of drift, where crime is possible but not inevitable~\cite{matza_delinquency_2018, paquet-clouston_extending_2025}. In such a drift state, cybercrime forum participation can be measured on a spectrum ranging from mere presence to explicit criminal engagement. In terms of criminal engagement, some users may decide to openly disclose criminal activities, while others may engage in crime, but choose not to reveal it publicly. Only the former are visible to researchers, whereas the latter remain difficult to detect, especially because of the ``grey area'' created due to the neutrality of IT~\cite{bijlenga_criminals_2018}. 
    
    Building on this idea, the present study assesses how often users explicitly post about crime in cybercrime forums. Specifically, this study \textbf{examines crime disclosure patterns in a major cybercrime forum} comprising over 3.5 million posts and nearly 300,000 users. This forum is particularly relevant because it is readily accessible and has been consistently conceptualized in prior research as a cybercrime or an underground forum~\cite{huang_hackerrank_2021, johnsen_identifying_2020, sun_understanding_2019, overdorf_under_2018, fang_analyzing_2019}. Crime disclosure is coded across three levels: benign, grey, and crime, to capture not only overt criminal activity, but also the size and scope of ambiguous or benign discussions. Moreover, given the structure of the cybercrime forum and the prevalence of short, sometimes out-of-context comments, the analysis focuses on initial thread posts. Following the study's objective, three research questions are developed. 

      \begin{description}
        \item[RQ 1:] What is the level of crime disclosure in the forum?       
        \item[RQ 2:] To what extent do forum participants transition between levels of crime disclosure?
        \item[RQ 3:] How does user's public posting behavior on crime disclosure relates to their private communication patterns on the forum?
    \end{description}

    First, we measure the extent of crime disclosure on the forum, then we investigate users' patterns of crime disclosure. Last, we assess whether these disclosure patterns are associated with the use of private communication channels. Our research questions are addressed using statistical analyses, including level-to-level analyses with Markov chains, alongside analyses of private messages using a logistic regression model and assortative mixing analysis.
    
    This study is the first to assess crime disclosure patterns in a large cybercrime forum and provides a scalable methodological approach to identifying and measuring these patterns. The study key takeaways can be summarized as follow:  

    \begin{itemize}
        \item Cybercrime forums foster a setting where crime disclosure is relatively normative, with one quarter of initial posts containing explicit crime-related content and more than one third of users disclosing criminal activity at least once in their initial posts.

        \item However, most participants also show restraint, with over two-thirds posting only benign or grey content in their initial posts. Within these posts, their crime disclosure tends to develop gradually, with most users starting with benign or grey content and escalating (if at all) toward explicit criminal disclosure one level at the time. 

        \item  Grey initial posts posts are particularly prominent in all analysis, indicating that many users avoid overt crime disclosure and anchor their activity in ambiguous content.
    \end{itemize}

    Our study highlights the value of scalable computational methods, indicating that Large Language Model (LLM)-based labelling and Markov chain modelling can reliably capture crime disclosure patterns while offering promising research on user behaviour and the evolution of cybercrime activity. Inspired by the study's design, law enforcement agencies can improve their detection and prevention efforts by more accurately identifying which users signal criminal involvement as opposed to benign or ambiguous activity. Moreover, comparing different cybercrime forums based on their level of crime disclosure could also help researchers and law enforcement agencies better prioritize which forums warrant closer monitoring or intervention.

\section{Literature Review}
    In this section, we present different streams of literature that analyse data from cybercrime forums. Next, we conceptualise cybercrime disclosure based on previous work and formulate objectives aligned with our research questions.

    \textbf{Cybercrime Forums as Offender Convergence Settings:} Cybercrime forums are offender convergence settings that serve market, social and learning purposes~\cite{leukfeldt_use_2017, soudijn_cybercrime_2012}. Since their inception, administrators have put mechanisms in place to foster social and economic relations among actors, including reputation scores and feedback systems~\cite{afroz_honor_2013, holt_examining_2013, dupont_darkode_2017, decary-hetu_reputation_2013}. These systems mitigate fraud risks and promote successful transactions~\cite{campobasso_cybercriminal_2023}. They have facilitated the emergence of a volume-based industry, in which actors tend to specialize in a single cybercrime task, such as creating phishing kits or stealing credentials, that can then be traded or monetized through cybercrime forums~\cite{moore_economics_2009, collier_cybercrime_2021, collier_cybercrime_2020, sembera_cybercrime_2021, paquet-clouston_motivations_2022, akyazi_measuring_2021}.
    
    However, when looking at the network dynamics of cybercrime forums, some studies have found that only a small proportion of users are highly connected, while most users maintain weak and sporadic ties~\cite{pete_social_2020, hughes_digital_2023}. Similarly, several studies have highlighted the highly uneven distribution of activity levels in cybercrime forums, where a small number of participants are extremely active while the vast majority exhibit only low levels of activity~\cite{hughes_digital_2023, CARRONARTHUR2014165, huang_hackerrank_2021, ruellan_identifying_2024}. This raises the question: what is the proportion of cybercrime forum users who are consistently involved in cybercrime? \citet{pastrana_characterizing_2018} mention in their study: ``Underground forums contain many thousands of active users, but the vast majority will be involved, at most, in minor levels of deviance. The number who engage in serious criminal activity is small'' (p.207)~\cite{pastrana_characterizing_2018}. These findings indicate that users who are present within cybercrime forums are not necessarily substantively involved in criminal activity. This highlights the need for a more nuanced approach that considers whether users maintain benign roles or gradually shift towards other levels of criminal engagement.
        
    Moreover, apart from facilitating the exchange of products, services, and knowledge related to cybercrime~\cite{leukfeldt_cybercriminal_2017, soudijn_cybercrime_2012}, these forums also host a wide range of topics, beyond cybercrime. Actors discuss various topics, such as technical problem solving, food or religion ~\cite{allodi_then_2016, dupont_darkode_2017, decary-hetu_reputation_2013}. These discussions are part of the social aspect of these forums, as individuals develop weak and strong relationships through these interactions. The criminal character of some discussions is, moreover, sometimes uncertain. This is due to the neutrality of IT: many tasks related to cybercrime, such as building websites or managing servers, are the same as any regular IT tasks~\cite{bijlenga_criminals_2018}. This grey area is further exacerbated by the increasing commoditization and specialization of the cybercrime industry, with tasks becoming more specific and repetitive \cite{moore_economics_2009, wegberg_plug_2018, collier_cybercrime_2020, collier_cybercrime_2021}. Together, this complicates the process of understanding and accurately measuring criminal participation in cybercrime forums.
        
    \textbf{Conceptualizing levels of Cybercrime Disclosure in Forums:} Given that most forum participants are only briefly active, that many benign topics are discussed, and that the ``grey zone'' of ambiguous activity is expanding, there is a need for a more nuanced way to understand participation in cybercrime forums. One interesting perspective comes from~\citet{paquet-clouston_extending_2025}, who draw on Matza's notion of drift~\cite{matza_delinquency_2018}.~\citet{matza_delinquency_2018} emphasized that there is no clear distinction between people involved in crime (delinquents) and those who are not (non-delinquents). Instead, people \textit{drift} in and out of criminal involvement, only temporarily integrating criminal values. Drifting is a condition of loosened social controls, it is a mental state in which the tie binding the self to legal moral expectations is weakened. In such a temporary state, crime is possible, but not necessarily inevitable~\cite{matza_delinquency_2018}.~\citet{goldsmith_digital_2015}  extended this idea to online contexts, suggesting that Internet features, such as pseudo-anonymity and low immediate consequences, can make such transitions easier.
    
    Conceptualizing cybercrime forum participation as akin to being in a state of drift underscores that forum presence alone does not directly equate to criminal engagement,  introducing an important layer of conceptual nuance. Moreover, while some users may openly disclose criminal activities in their posts, making their involvement observable, others may engage in crime but choose not to reveal it publicly. The latter group cannot easily be quantified, and identifying them is further complicated by the neutrality of IT~\cite{bijlenga_criminals_2018}. Nevertheless, this theoretical lens helps differentiate between mere participation and explicit criminal disclosure on these forums. Building on this, the present study examines the extent to which users explicitly reveal criminal activities in their posts, while acknowledging that such disclosures represent only a subset of actual criminal engagement.
    
    Specifically, \textbf{this study examines crime disclosure patterns within a major cybercrime forum}. Participants' initial posts are classified into three levels of disclosure: (1) benign, (2) grey, and (3) crime-related. This approach sheds light on how often users openly disclose criminal activity in cybercrime forums, while also allowing an assessment of the size and scope of both benign and ``grey'' (i.e., ambiguous) activities.  Given the structure of the cybercrime forum and the prevalence of short, sometimes out-of-context comments, the whole analysis focuses on initial thread posts. Three objectives are derived from the research questions presented above. They are as follows:
    
    \textbf{Obj.1}: Given these labels, we measure the level of crime exposure present in initial posts. This provides an initial overview of how much explicit criminal activity is publicly revealed. Given that the platform is labelled a cybercrime forum, we expect a high level of crime exposure. However, prior research also suggests that crime exposure in such forums is often concentrated within a relatively small subset of users~\cite{pastrana_characterizing_2018}.
    
    \textbf{Obj.2}: Second, using Markov chains, the study analyzes how participants switch between the three disclosure levels. Because cybercrime forums function as offender convergence settings, users may follow different exposure patterns. Some may openly reveal their criminal activity from the outset. Others may be influenced by the criminogenic environment and display a gradual progression in exposure (benign to grey to crime). Finally, some may never disclose any criminal activity (stay in grey or benign), either because they are not involved or because they wish to avoid self-incrimination for security reasons.
    
    \textbf{Obj.3}: Finally, the study assesses how public posting behavior in crime disclosure relates to private communications. Private messages may serve as an alternative channel for individuals who deliberately restrain their public exposure, yet still wish to discuss criminal activities. 

\section{Methodology}

    Addressing these objectives requires several complementary analytical steps. First, initial thread posts are classified into disclosure levels using large language models (LLMs). Second, because activity levels in cybercrime forums and online forums, are highly uneven, with a small number of highly active users and a large majority of low-activity participants~\cite{hughes_digital_2023, CARRONARTHUR2014165, huang_hackerrank_2021, ruellan_identifying_2024}, we perform a clustering analysis to distinguish the general population from the most active users.

    Analytically, the first objective is addressed through descriptive statistics and visualizations of disclosure patterns. The second objective relies on Markov-chain modeling to examine transitions between disclosure levels. Finally, the third objective is achieved through a logistic regression and an assortative-mixing analysis, made possible through the graph representation of private message relationships between users.

    \subsection{Dataset}
        For this study, the leaked and publicly available database of the cybercrime forum ``nulled.io'' is used. This database contains public and private messages of forum users, as well as their user account information (under their pseudonyms). The data ranges from January 2015 to May 2016. There are around 3.5 million posts and $\sim$400,000 private messages in the database. The content is predominantly in English, with only limited exceptions. Within the forum, users could start new discussion threads or comment existing ones in multiple sub-forums with names such as `Introductions', `League of Legends', `Malicious Tools', or `Cracking Support'. The start of a new thread is referred as initial thread post and other users posts/responses as comments.
        
        The leaked nulled.io dataset has been studied by other researchers for different purposes, such as identifying proficient hackers~\cite{huang_hackerrank_2021, johnsen_identifying_2020}, predicting private interactions ~\cite{sun_understanding_2019, overdorf_under_2018}, and identifying data breaches on cybercrime forums ~\cite{fang_analyzing_2019}.  We selected this forum for three main reasons. First, it was readily accessible. Second, prior research consistently conceptualizes it as a cybercrime or underground forum~\cite{huang_hackerrank_2021, johnsen_identifying_2020, sun_understanding_2019, overdorf_under_2018, fang_analyzing_2019}. Third, the database includes private message information, which is necessary for addressing the third objective.
 
\begin{figure}[htb]
            \centering
            \includegraphics[width=.95\linewidth]{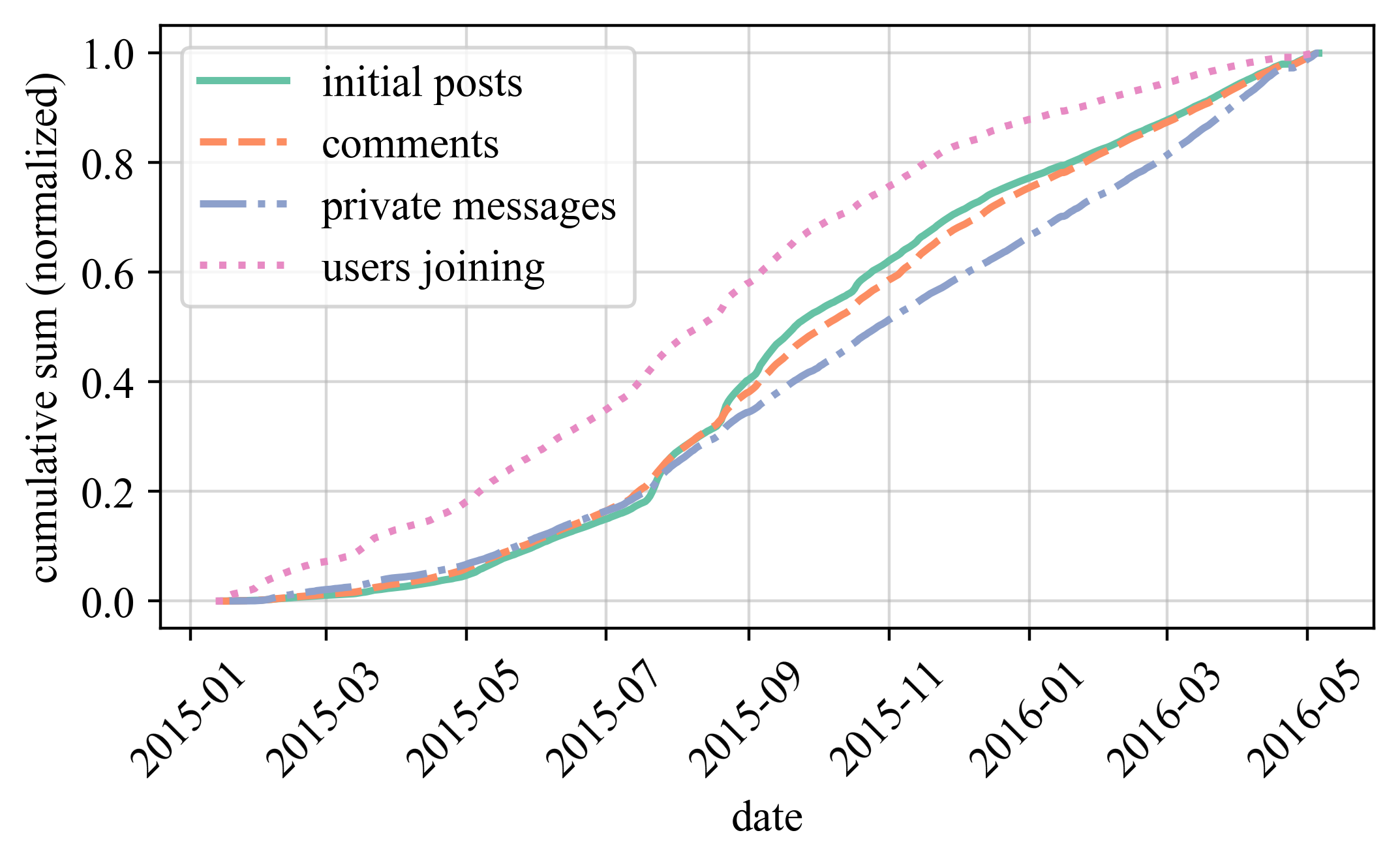}
            \caption{The cumulative sum of initial posts, comments, private messages, and the number of users joining the forum.}
            \label{fig:dataset_descript_cumsum}
        \end{figure}

        Figure \ref{fig:dataset_descript_cumsum} shows the cumulative normalized number of users joining the forum, as well as initial posts, comments, and private messages. The figure reveals a generally steady increase, with a pronounced rise during the summer months from July to September 2015. Table \ref{tab:dataset_overview} complements this by summarizing the number of users joining the forum, the distribution of initial posts across disclosure levels, and the number of users who exchanged private messages.
    
        \begin{table}[htb]
            \small
            \centering
             \caption{The Table provides an overview of user activity on the forum.}
            \label{tab:dataset_overview}
            \begin{tabular}{lrr}
                \toprule
                Description & 2015 & 2016 (Jan-May) \\
                \midrule
                Newly registered users & 480,380 & 118,705 \\
                Users commenting initial posts & 232,130 & 93,529\\
                Users writing private messages (UPM) & 23,968 & 9,042\\
                Users writing initial posts (UIP) & 27,003 & 8,319\\
                Intersection UPM and UIP & 12,238& 4,027\\ 
                Private messages  &336,559 & 171,155\\
                Unclear initial posts &3,820 &661 \\
                Benign initial posts & 30,674& 5,651\\
                Grey initial posts & 35,637& 13,627\\
                Crime-related initial posts& 23,138&7,801 \\
                \bottomrule
    
            \end{tabular}
        \end{table}
        
        Despite being published several years ago, the nulled.io dataset offers a unique opportunity to retrospectively examine how users disclose their level of crime engagement. 

    \subsection{Crime disclosure labeling using LLM}\label{sec:labeling}
        To examine crime exposure patterns, we categorized initial thread posts into benign, grey or crime-related using a LLM. We chose to apply an LLM as a way to automate the labelling process. Recent work has shown that LLMs can successfully classify content from cybercrime forums~\cite{moreno-vera_inferring_2023, clairoux-trepanier_use_2024}. Nevertheless, this data processing was particularly intensive, as described in more detail below.
    
        To prepare the dataset for the labelling tasks and subsequent analyses, we first processed and cleaned the data. This involved removing HTML tags and quoted text, as well as removing URLs and email addresses. The cleaning procedures relied on functions provided by the Python library Textacy~\cite{textacy_2025}. We also excluded users whose account-creation date was unknown or whose posts had timestamps preceeding their join date. This affected 13 users and 2,421 posts. The final dataset comprises 3,492,501 posts, 121,032 threads, and 299,682 users who contributed initial posts and/or comments to the forum.

        \begin{figure}[hbt]
            \centering
            \footnotesize
            \begin{tikzpicture}
                \node (first) at (0,0) [header=Process Posts \& Messages, standard] {Clean public posts. Remove HTML tags, quoted text, unicode, and replace emails and URLs with placeholders \texttt{\_EMAIL\_} and \texttt{\_URL\_}. \\ Filter Private Messages. Remove 286,643 system messages.};
                \node (second) at (4.6,0) [header=Thread Comment Summary, standard] {Use LLM to summarize all comments in a thread. LLM Input: \begin{itemize}
                    \item Thread title
                    \item Sub-Forum name
                    \item All posts of the thread
                \end{itemize}};
                \node (third) at (4.6,-4.5) [header=Crime Disclosure Labelling, standard] {Use LLM to label the initial post of all threads. LLM Input: \begin{itemize}
                    \item Thread title
                    \item Sub-Forum name
                    \item Summary of comments
                    \item Content of the initial post
                \end{itemize}};
                \node (forth) at (0,-4.5) [header=Activity Clustering,standard,] {K-Means clustering with variables:
                \begin{itemize}
                    \item Number of calendar weeks in which a users writes an initial post
                    \item Maximum number of weeks between writing initial posts
                    \item Number of related initial posts per disclosure level (four variables)
                \end{itemize}};
                \node (fifth) at (2.3,-8.45) [header=Analysis, wide] {
                \vspace{-2em}
                \hspace{-1em}
                \begin{tabular}{p{0.215\textwidth} p{0.215\textwidth} p{0.215\textwidth} p{0.215\textwidth}}
                \textbf{Descriptive analysis,} showing the distribution of crime disclosure levels. $\rightarrow$ Obj. \ding{172} &
                \textbf{Markov-Chains,} showing the transitions between disclosure levels per cluster. \ \ \ \ $\rightarrow$ Obj. \ding{173}&
                \textbf{Statistical Analysis,} assessing if disclosure level relates to private messaging with log. reg. \phantom{XXXX} $\rightarrow$ Obj. \ding{174}&
                \textbf{Assortative Mixing,} assessing whether users messaged others with similar disclosure profiles.\phantom{XXXX} $\rightarrow$ Obj. \ding{174}
                \end{tabular}
                };
                
                \path [connector] (first) -- (second);
                \path [connector] (second) -- (4.6,-2.46);
                \path [connector] (third) -- (forth);
                \path [connector] (forth) -- (0,-6.95);
    
            \end{tikzpicture}
            \caption{Overview of the processing steps of our methodology, starting with four preprocessing steps followed by three analysis methods.}
            \label{fig:overview_prepro}
        \end{figure}
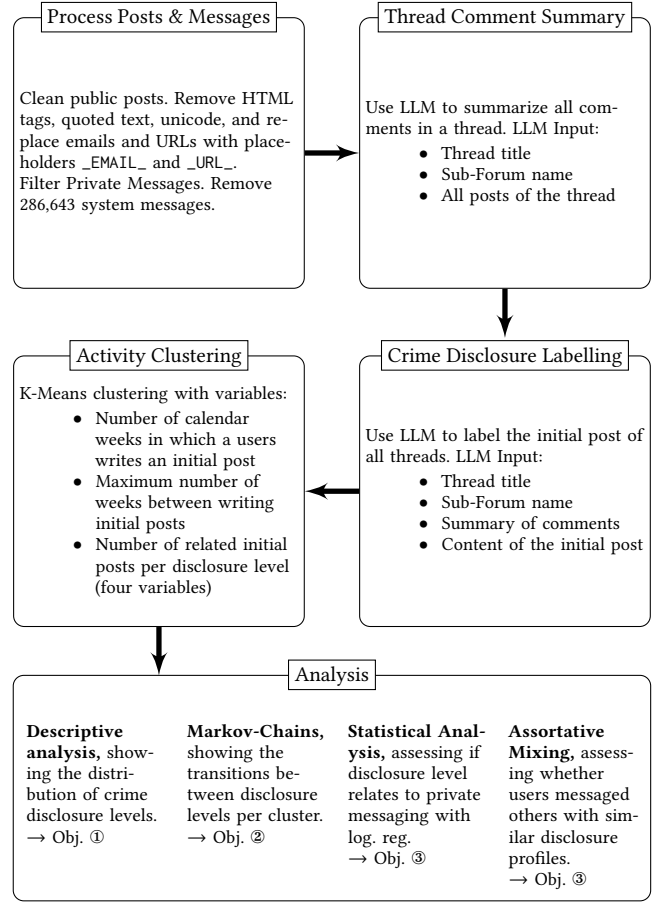
        
        Each initial post was classified according to its degree of crime disclosure. The levels were defined as:
    
        \textbf{Crime Related:} Posts that directly involve the promotion, sale, solicitation, or exchange of illegal goods/services or criminal knowledge, or that clearly violate national/international laws.
        
        \textbf{Grey:} Posts that do not clearly break the law, but may facilitate or enable criminal activity, or whose legality and intent are context-dependent. These can fall into ethical or legal grey zones.
        
        \textbf{Benign:} Posts that do not involve any form of illegal or ethically questionable activity nor enable or facilitate such actions.
        
        A fourth category, \textbf{Unclear}, was added to capture posts that did not contain enough information to assign any of these three labels. For the labelling process, we used the OpenAI GPT-4.1 model family. The classification was performed through OpenAI’s batch API, processing up to 30,000 posts per batch, and we relied on the largest available GPT-4.1 model at the time to ensure the highest possible quality of results.
        
        To evaluate the LLM's performance at categorizing posts, we created a ``ground-truth'' dataset. Specifically, four researchers manually labelled 100 posts and met several times to resolve posts with substantial disagreement. Then, this dataset was used for calculating performance metrics. Using the prompts shown in Figure~\ref{app:system_prompt} and Figure~\ref{app:user_prompt} (see Appendix), the LLM achieved an accuracy of 83.0, an $F_1$ score of 83.7, and a precision of 84.4. To assess the consistency of the LLM, we also drew a random sample of 100 threads and repeated the labeling procedure ten times. The outputs showed an average pairwise agreement of 95\%, with a minimum of 92\%. Such results illustrate that the LLM's coding output was consistent. These results were deemed sufficiently robust to support the use of the LLM-generated categories as the analytical units in the study.
        Table \ref{tab:ex_init_posts} provides examples of initial posts for each disclosure level.
       \begin{table}[htb]
            \small
            \centering
             \caption{The Table shows examples of initial posts for the disclosure labels, benign, grey, and crime-related. Some characters have been removed for better readability.}
            \label{tab:ex_init_posts}
            \begin{tabular}{p{0.75\linewidth} r}
                \toprule
                Initial Post & Label \\
                \midrule
                Please, I'm working into colors and edition in Photoshop CC 2015. Need some feedbacks, I'll post the before and after! I'm trying to do it faster , so it's not anything especial or professional.[...] BEFORE: CLICK HEREAFTER: CLICK HERE [...] The quality is bad due to image res. & benign\\
                So here's an off-topic question, what's your favorite movie or movie's? I personally like the new Jurassic World, Zombie-land, and  the R.E.D series.What about you? & benign\\
                \_URL\_This is only really useful for Origin accounts as there are no accounts for the other sections. & grey\\
                A· Inactive means inactive for 3+ months [..] NO PERSONAL ACCOUNTS- Only EUNE Accounts [...] Only level 30 accounts with 60+ Champions   Contact me via PM giving full info of the accounts & grey\\
                I got this source code for a java rat and idk how to compile it to use it can someone help? & crime-related \\
                hey guys :) That's my share : ALL VERIFIED, CRACKED ACCOUNT  \_URL\_ +rep if you want more :) & crime- related \\
                \bottomrule
            \end{tabular}
        \end{table}

    \subsection{Clustering analysis}

        Once all initial posts were labeled, we applied a clustering analysis to distinguish highly active users from the rest of the forum population, a pattern well-documented in cybercrime and online forums~\cite{hughes_digital_2023, CARRONARTHUR2014165, huang_hackerrank_2021, ruellan_identifying_2024}. This separation allowed us to compare results across these two clusters and, in some cases, uncover distinct behavioral patterns that would otherwise remain obscured in the aggregated data. We first selected all users who started, at least once, a thread in the cybercrime forum (N=33,127 thread initiators). Then, to divide this sub-population, the clustering analysis used six variables:
        \begin{itemize}
            \item the number of active weeks (weeks in which a user posted at least one initial post),
            \item the number of inactive weeks (highest number of weeks beginning with the join date in which the user made no initial posts),
            \item the total number of crime-related initial posts,
            \item the total number of grey initial posts,
            \item the total number of benign initial posts, and
            \item the total number of unclear posts.
        \end{itemize}

        We applied k-means clustering and used the silhouette method to determine the optimal number of clusters. A two-cluster solution produced the highest silhouette score (0.783), closely followed by a three-cluster solution (0.771). We selected the two clusters solution because of the slightly higher score and because it provided a clear distinction between highly active users and the rest of the population.
        
        Table~\ref{tab:descriptives_groups} presents the distribution of the variables across the two clusters. Due to the uneven distributions and non-normality (verified using Q-Q plots), we performed Mann-Whitney U(MWU) tests to assess the degree to which each cluster differed based on the variables. The MWU test checks whether the distribution of the variables for both clusters are the same or not. To account for the imbalance in the cluster sizes, we also performed bootstrapping, which confirmed the results of the MWU test. All tests were significant. The first cluster contains 30,242 users (91\% of thread initiators), while the second cluster consists of 2,885 users (9\% of thread initiators). The descriptive statistics indicate that the first cluster represents the general population of low-activity users: for every variable, both the mean and median values are lower than those of the second cluster, which corresponds to the forum highly active users.
        
        ~\textbf{In sum, Figure~\ref{fig:overview_prepro} provides an overview of the full data-processing, labelling and clustering workflow.} 
    
        \begin{table}[htb]
            \centering
            \footnotesize
             \caption{The table presents descriptive statistics for the variables underlying the two user clusters, Cluster 1 (C1) and Cluster 2 (C2). Reported measures include the mean, standard deviation, and median, along with the Mann–Whitney U test results and their statistical significance (***).}
            \label{tab:descriptives_groups}
            \begin{tabular}{lccccc}
                \toprule
                 & \multicolumn{2}{c}{Mean (std)} & \multicolumn{2}{c}{Median} &  \\
                Variable & C1 & C2 &C1 & C2 &{MWU Stat} \\
                \midrule
                 active weeks&1.35 (0.98)&6.09 (5.70)&1.0&4.0&$4.80 \cdot  10^6$*** \\
                 max inactive weeks &0.54 (1.66)&16.56 (9.24)&0.0&15.0&$1.32 \cdot 10^6$***\\
                 crime ip count &0.59 (1.34)&4.53 (10.23)&0.0&2.0&$2.00 \cdot 10^7$*** \\
                 grey ip count &0.90 (1.61)&7.59 (14.78)&0.0&3.0&$1.51 \cdot 10^7$*** \\
                 benign ip count &0.92 (1.71)&2.90 (6.86)&1.0&1.0&$3.19 \cdot 10^7$***\\
                 unclear ip count &0.10 (0.48)&0.48 (2.21)&0.0&0.0&$3.75 \cdot 10^7$*** \\
                 \bottomrule
                 \multicolumn{4}{l}{\rule{0pt}{1.2\normalbaselineskip}
                 \textsuperscript{***}$p \leq 0.001$, 
                 \textsuperscript{**}$p \leq 0.01$, 
                 \textsuperscript{*}$p \leq 0.05$}
            \end{tabular}
        \end{table}

    \subsection{Measuring the level of crime exposure in the forum (Obj.1)}
        To address the first objective on measuring the level of crime exposure in the forum, we first computed \textbf{descriptive statistics} on the LLM-generated labels. We used a \textbf{Venn diagram} (see Figure~\ref{fig:venn_user_disclosure}) to show overlap between different categories of users based on their level of crime exposure (e.g., individuals who post only crime-related content or those whose initial posts span all categories). These visual and statistical summaries allowed us to clearly illustrate the diversity of crime disclosure patterns within the forum.

    \subsection{Assessing how participants switch between crime disclosure levels (Obj.2)}
        To address the second objective, we employed \textbf{Markov chain} analysis. A Markov chain is a stochastic process defined by transitions between a finite or countable set of states, where the probability of moving to the next state depends solely on the current state and not on the preceding sequence of events \cite[p.~16]{pardoux_markov_2008}. 
        
        Originally developed in the early 20th century, Markov chains have since been widely used to analyse sequential data across numerous domains \cite[p.~464]{grinstead_introduction_1998, Vermeer_Trilling_2020}. In recent years, researchers have increasingly applied Markov chains and related models to study online user behavior, for example, to characterize interactions within mobile applications \cite{liu_characterizing_2019}, to analyse shifts in posting intent in mental health-related online communities \cite{morini_participant_2025}, and to trace users’ news reading trajectories \cite{Vermeer_Trilling_2020}. Inspired by this line of work, we leveraged stationary Markov chains to assess how forum participants transition between different levels of crime disclosure. 
        
        The analysis produced a transition-probability matrix, which we visualized as a map representing the likelihood that users move between disclosure levels. We also included two additional states: a start state (representing when a user first joins the forum) and an end state (representing when a user stops posting initial threads).

    \subsection{Assessing how public posting behavior relates to private communications (Obj.3)}
        To address the third objective, we conducted two analyses: a logistic regression model \cite{noauthor_introduction_2018} and an assortative mixing analysis. The \textbf{logistic regression} estimated the probability that a user sent at least one private message as a function of their number of crime-related, grey, benign, and unclear posts, as well as the cluster they belonged to. Due to skewed distributions, we used log-transformed counts, defined as $C_i = \log(\text{CrimePosts}_i + 1)$, $G_i = \log(\text{GreyPosts}_i + 1)$, $B_i = \log(\text{BenignPosts}_i + 1)$, $U_i = \log(\text{UnclearPosts}_i + 1)$, and $D_i = \log(\text{DaysToEnd}_i + 1)$. The model was specified as:
        
        \begin{equation}
        \begin{split}
        \text{logit}\left( P(\text{PM}_i = 1) \right)
        &= \beta_0 
        + \beta_1 C_i
        + \beta_2 G_i 
        + \beta_3 B_i \\
         &\quad+ \beta_4 U_i
        + \beta_5 D_i
        + \beta_6 \,\text{Cluster}_i
        + \epsilon_i.
        \end{split}
        \end{equation}
        
        This model enabled us to identify which forms of public posting behavior were predictive of private messaging activity, thereby indicating how public crime exposure relates to communication in private channels.
        
        For the second analysis, we created an undirected graph representing private-message exchanges between users who posted at least one initial thread. For each user, we also identified their dominant crime-disclosure category. We then applied an \textbf{assortative mixing analysis} on the graph to examine whether users preferentially exchanged private messages with others who shared similar dominant disclosure profiles. The results are expressed as a mixing matrix summarizing the relative frequency (in proportions) of connections between dominant disclosure profiles. This analysis allowed us to examine whether private communication tends to occur between users with similar disclosure profiles.

\section{Results}\label{sec:results}

    In the present study we analyse data of the nulled.io forum to address the research questions about crime disclosure. In the following we present the results of our analyses where each subsection relates to one research question/objective.

    \subsection{Crime disclosure in initial thread posts (Obj.1)}

        To begin, it is worth noting that only a minority (approximately 11\%, N=32,127) of users on the forum started threads. The remaining (89\%, N=267,555) contributed solely through follow-up comments or private messages. Initial thread posts were classified as benign, grey, crime, or unclear to assess the extent of crime disclosure in the forum. 
        
        As shown in Table~\ref{tab:label_overview_severity}, initial threads were categorized as grey (40.71\%), benign (30.01\%) and crime-related (25.57\%), with only a small proportion labelled as unclear (3.70\%). Follow-up comments were most frequent on threads with a crime-related initial post (46.90\%), followed by grey initial posts (41.37\%) and benign initial posts (10.14\%). This indicates lower user engagement in threads initiated with benign posts compared to those initiated with crime-related content.
    
         \begin{table}[hbt]
            \centering
            \small
            \caption{Overview of the Crime Disclosure Labelling. The comments to a thread (initial post) have the same label as the initial post. The value in the brackets shows the corresponding percentage or standard deviation (last column on the right).}
            \label{tab:label_overview_severity}
            \begin{tabular}{@{}lrrr@{}}
                \toprule
                Label &\begin{tabular}[c]{@{}l@{}}\#~Initial Posts \end{tabular}& \begin{tabular}[c]{@{}l@{}}\#~Comments\\to initial posts \end{tabular}& \begin{tabular}[c]{@{}l@{}}Mean \#~comments \\ per thread\end{tabular} \\
                \midrule
                Grey & 49,278 (40.71\%) & 1,394,872 (41.37\%) & 32.86 (149.91)\\
                Benign & 36,328 (30.01\%) & 342,047 (10.14\%) & 12.08 (40.23)\\
                Crime-related & 30,944 (25.57\%)& 1,581,104 (46.90\%) & 58.00 (455.03)\\
                Unclear &4,481 (3.70\%) &53,405 (1.58\%) & 15.23 (46.85)\\
                \bottomrule
                
            \end{tabular}
        \end{table}

\begin{figure}[htb]
                \centering
                \includegraphics[width=0.7\linewidth]{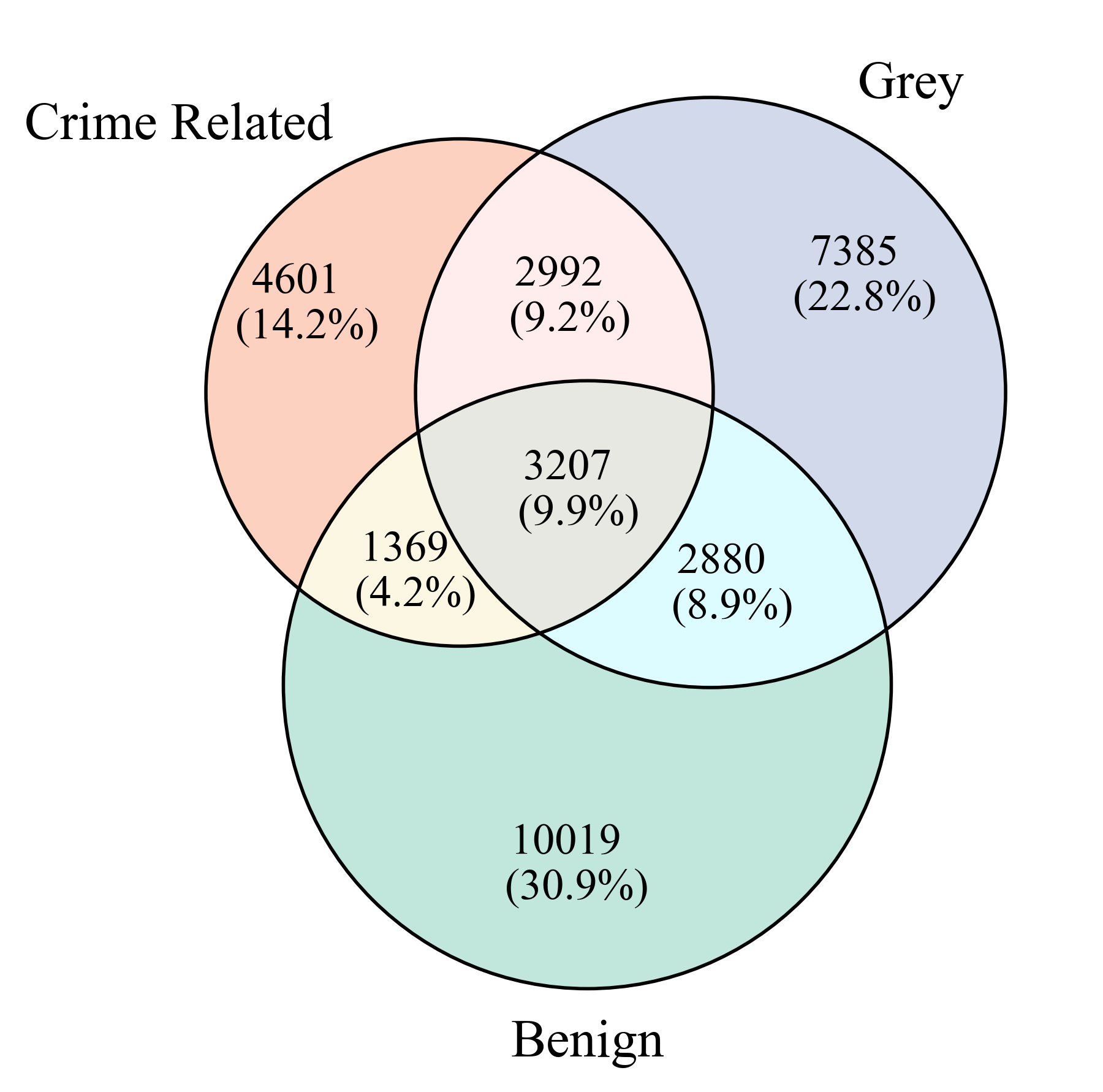}
                \caption{Diagram showing the the number of users who write initial posts per disclosure label and the overlap between users who write initial posts with different crime disclosure levels.}
                \label{fig:venn_user_disclosure}
        \end{figure}

        The Venn diagram in Figure~\ref{fig:venn_user_disclosure} shows that nearly one-third of users (30.9\%) write only benign initial posts while 22.8\% write exclusively grey content, and 14.4\% write only crime-related initial posts. Note that we removed unclear posts from this visualization. Although the proportion of users who post exclusively crime-related content is relatively small (14.2\%), when including all users who initiated at least one crime-related post, the share increases to 37.5\%. Thus, slightly more than one-third of users publicly expose some form of criminal activity. On the other hand, 62.6\% of users wrote only benign or grey initial posts.

        \noindent \textbf{Key takeaway: More than a third of users disclosed, at least once, crime-related content in their initial posts. On the other hand, about two-third of users wrote only benign or grey initial posts.}

    \subsection{Crime disclosure level transitions (Obj.2)}
\begin{figure}[htb]
                \centering
                \includegraphics[width=0.95\linewidth]{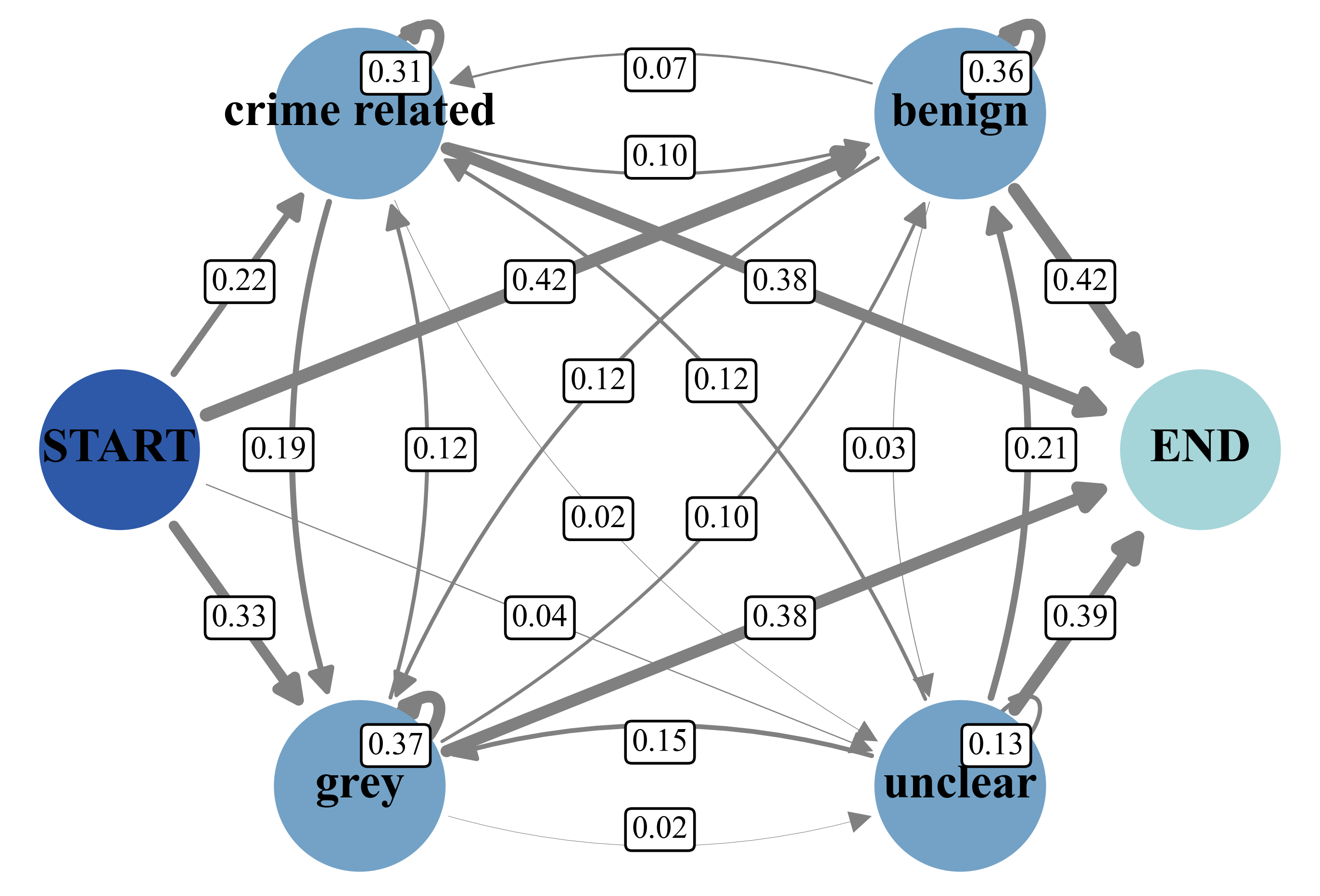}
                \caption{Markov chain visualization for Cluster 1. Circles represent disclosure levels of initial posts, while arrows indicate transition probabilities between levels.}
                \label{fig:markoc_cl0}
        \end{figure}

\begin{figure}[htb]
            \centering
            \includegraphics[width=0.95\linewidth]{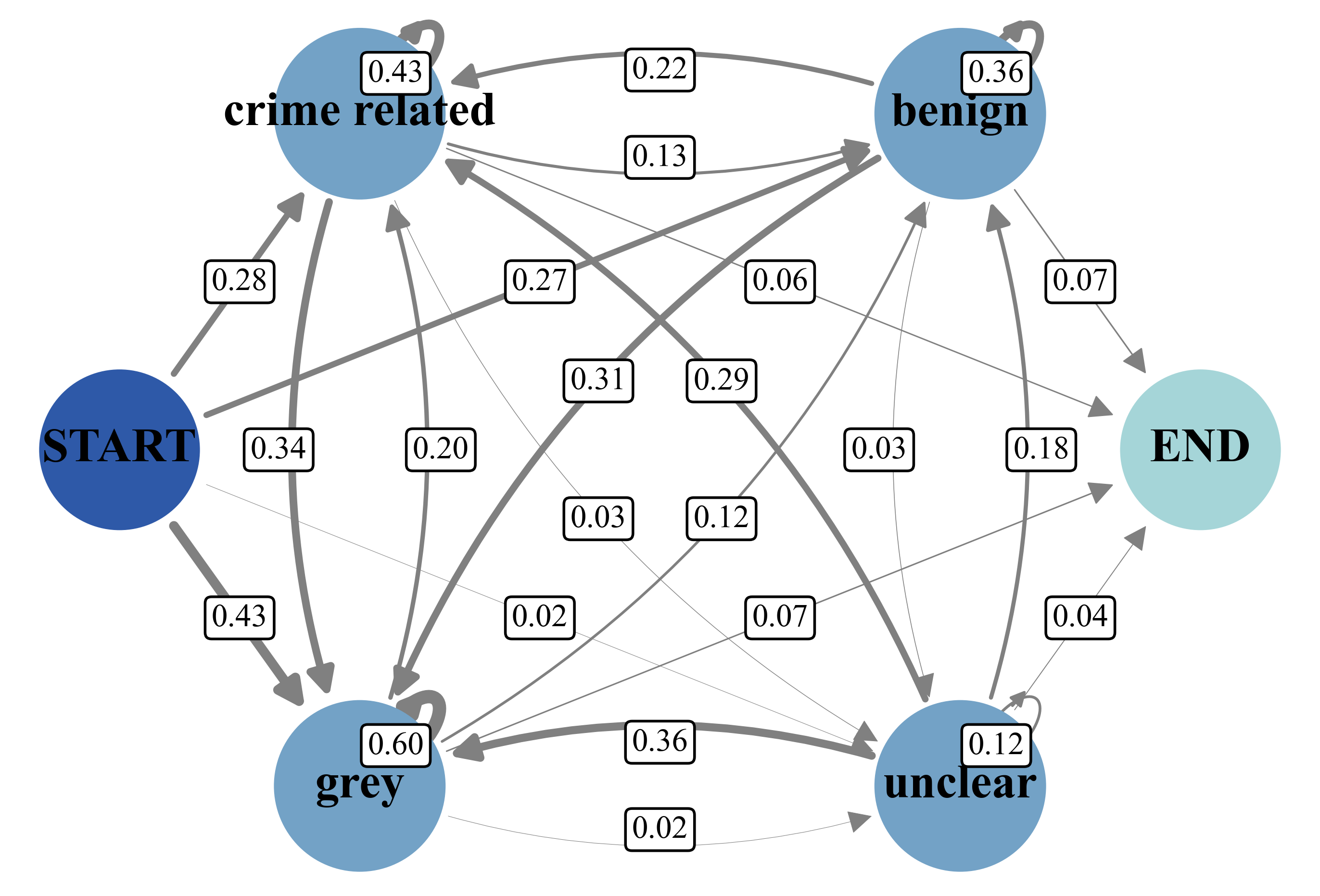}
            \caption{Markov chain visualization for Cluster 2. Circles represent disclosure levels of initial posts, while arrows indicate transition probabilities between levels.}
            \label{fig:markoc_cl1}
        \end{figure}

        To examine how participants transitioned between crime-disclosure levels, we computed Markov chains separately for the two clusters. This approach allowed us to assess disclosure patterns in the general population (Cluster 1, Figure \ref{fig:markoc_cl0}) and among highly active users (Cluster 2, Figure \ref{fig:markoc_cl1}).
        
        The Markov chains show that users in the general population (Cluster 1) are most likely to begin their public activity with a benign thread (0.42), whereas highly active users (Cluster 2) most commonly start with a grey post (0.43). Moreover, in both clusters, fewer than one third of users began their activity with a crime-related thread.
        Another clear pattern in both Markov chains is the prominence of self-loop probabilities, indicating a strong persistence within the same crime-disclosure level. In other words, users tend to remain in the same disclosure state from one initial post to the next rather than transition to a different level. This tendency is especially pronounced among highly active users in Cluster 2, where the probability of writing a subsequent grey post after a previous grey post reaches 60\%.
        Moreover, in both Markov chains, transitions typically occur between adjacent disclosure levels rather than jumping two levels at once. For example, for both clusters, users are more likely to write a grey initial post after a benign one than to move directly from benign to crime-related content.
        Toward the end of the activity sequence, the high probability of transitioning to the end state among the general population (Cluster 1) reflects the shorter activity spans of these users. Many in this cluster write only one or two initial posts before ceasing to contribute further.
        
        \noindent \textbf{Key takeaway: Users are unlikely to start with crime-related content, tend to remain within the same disclosure level due to strong self-loops, and when they do shift, transitions occur mainly between adjacent levels rather than through abrupt jumps.}
    
    \subsection{Private messaging and crime disclosure (Obj.3)}
    
        Lastly, to assess how public posting behavior relates to private communications, a logistic regression and an assortative mixing analysis were computed. 
        
        The logistic regression estimates the probability that a user sends at least one private message; the results of the model are presented in Table \ref{tab:glm_results}.
        Writing more initial posts is significantly associated with a higher likelihood of having sent at least one private message, although the magnitude of this effect varies by disclosure types. Users who write more grey posts are 3.68 times more likely to have sent a private message (CI = 3.50–3.86, p $<$ 0.000), while those who write more crime-related posts are 2.64 times more likely to do so (CI = 2.50–2.79, p $<$ 0.000). In comparison, writing more benign posts increases the likelihood of having sent a private message by 1.42 times (CI = 1.35–1.49, p $<$ 0.000), and unclear posts show no significant effect. Highly active users (Cluster 2) are also 2.25 times more likely to have sent a private message than the general population (Cluster 1; CI = 1.98–2.56, p $<$ 0.000), and users with longer activity spans similarly exhibit higher odds (OR = 1.23, CI = 1.18–1.27, p $<$ 0.000). The McFadden pseudo $R^2$ of 0.164 indicates an adequate model fit, while also suggesting that additional variables may contribute to explaining private-messaging behavior.

        \begin{table}
            \small
            \caption{Results of the logistic regression model examining the likelihood of sending a private message (0 = no, 1 = yes). Among users who wrote an initial post (n = 33,127), 46\% (n = 15,267) wrote at least one private message. The model has a $R^2_{McF}$ of $0.164$.}
            \label{tab:glm_results}
            \begin{tabular}{l r l r r}
            \toprule
            {Variable} & {Estimate} & {Std. Error} &  {OR} & {95\% CI} \\
            
            \midrule
            Intercept & $-2.45$ *** & $0.10$  & $0.09$ & $[0.07, 0.11]$ \\
            UnclearPosts & $-0.04$ \phantom{***} & $0.06$ & $0.96$ & $[0.86, 1.08]$ \\
            GreyPosts & $1.30$ ***& $0.03$  & $3.68$ & $[3.50, 3.86]$ \\
            BenignPosts & $0.35$ ***& $0.02$  & $1.42$ & $[1.35, 1.49]$ \\
            CrimePosts & $0.97$ ***& $0.03$  & $2.64$ & $[2.50, 2.79]$ \\
            Cluster 1 & $0.81$ ***& $0.07$ &  $2.25$ & $[1.98, 2.56]$ \\
            DaysToEnd & $0.21$ ***& $0.02$  & $1.23$ & $[1.18, 1.27]$ \\
            \bottomrule
            \multicolumn{4}{l}{\rule{0pt}{1.2\normalbaselineskip}
            \textsuperscript{***}$p \leq 0.001$, 
            \textsuperscript{**}$p \leq 0.01$, 
            \textsuperscript{*}$p \leq 0.05$}
            \end{tabular}
        \end{table}

        \begin{table}
            \footnotesize
            \caption{Matrix displaying the results of the assortative mixing analysis linking private-messaging ties to dominant public crime-disclosure profiles. Rows and columns represent users who predominantly write initial posts at each specific disclosure level.}
            \label{tab:Assortative_mixing}
            \begin{tabular}{lccr}
            \toprule
                \text{}& \begin{tabular}[c]{@{}c@{}}\text{Benign-dom. users} \\ $n=3774$ \end{tabular}& \begin{tabular}[c]{@{}c@{}}Crime-dom. users \\ $n=4664$ \end{tabular} & \begin{tabular}[c]{@{}r@{}} Grey-dom. users \\ $n=7069$ \end{tabular} \\
                \midrule
                Benign-dom. users &0.23 &0.27& 0.50\\
                Crime-dom. users & 0.14 & 0.34 & 0.51 \\
                Grey-dom. users & 0.12 & 0.25 &0.63\\
                \bottomrule
                \end{tabular}
        \end{table}

        The second analysis builds on an undirected graph representing private-message interactions between users in the forum. For each user, we first identified their dominant crime-disclosure category and then applied an assortative mixing analysis to assess whether users tend to communicate privately with others who share similar disclosure profiles. Table~\ref{tab:Assortative_mixing} presents the resulting mixing matrix, summarizing the extent to which private-message exchanges occur within or across disclosure categories.
        
        Among users who predominantly post benign content in public, half of their private-message (50\%) go to grey-dominant users, while only 23\% go to benign-dominant users and 27\% to crime-dominant users, as shown in Table~\ref{tab:Assortative_mixing}. Users who mainly publicly post crime content have 14\% of their private messages going to benign-dominant users, 34\% to crime-dominant users, and 51\% to grey-dominant users. Users who mainly publicly post grey content have 12\% of their private messages going to benign-dominant users, 25\% to crime-dominant users, and 63\% to grey-dominant users.
        
        Across all groups, grey-dominant users are the primary private-message partners in the forum. Regardless of whether individuals mainly post benign, crime, or grey content publicly, they direct the largest share of their private interactions toward grey-dominant users. Grey-dominant users also show strong internal homophily, with nearly two-thirds of their private messages exchanged with other grey-dominant users. In contrast, crime-dominant users do not preferentially contact crime users; instead, they interact more frequently with grey users. 
        
        \noindent \textbf{Key takeaway: Grey-level disclosure is found to be the strongest driver of private messaging: users who write more grey posts are most likely to have private exchanges and the majority of private messages are directed toward grey-dominant users.}
    
\section{Discussion}\label{sec:discussion}

    This study is the first to assess crime disclosure patterns in a large cybercrime forum and provides a scalable methodological approach to identifying and measuring these patterns. We discuss the implications of the results below. 

    The results suggest that the forum fosters an environment where posting crime is relatively normative: one quarter of initial posts contain explicit crime-related content, which generates substantial user engagement (namely, follow-up comments), and more than one third of users disclose criminal activity at least once in their first posts.  While we conceptualized cybercrime forum participation as akin to being in a state of drift, where crime is possible but not inevitable~\cite{matza_delinquency_2018, paquet-clouston_extending_2025}, it is interesting to note that a significant proportion of the population did go beyond the drift state and explicitly disclosed crime, even if it is not a de-facto behavior. The proportion is, moreover, expected to increase when accounting for those involved in crime who refrained for disclosing it in on the forum. Hence, as expected, cybercrime forums act as offender convergence settings~\cite{soudijn_cybercrime_2012, leukfeldt_use_2017}, where criminal knowledge, networks, and opportunities are openly discussed. The results also align with~\citet{goldsmith_digital_2015}'s digital drift concept, which hypothesizes that individuals can move into and out of offending behaviors easily in online environments. 
    
    While \citet{pastrana_characterizing_2018} found that most forum users engage only in minor deviance or no crime at all, this study did not examine the specific types of crimes disclosed by forum participants. Consequently, some of the crime reported in our data may also fall into the category of minor deviance identified by \citet{pastrana_characterizing_2018}. Further studies should explore the nature and severity of disclosed criminal activity in greater details.
    
    On the other hand, although many individuals eventually disclose criminal activity, the results also reveal an interesting pattern of restraint that, in some cases, leads to explicit crime disclosure. Specifically, more than two-thirds of users post only benign or grey content, and they are generally unlikely to begin their participation with crime-related posts. Moreover, the findings show that users tend to remain at the same level of crime disclosure or escalate by only a single level, suggesting that disclosure unfolds gradually rather than through abrupt shifts. This pattern indicates that users' decisions to disclose criminal activity may follow an incremental process, both for the general population and highly active users. Future research should therefore investigate the factors that influence individuals to move towards explicit crime disclosure. 

    This restraint is, moreover, consistently reflected in the prominence of grey initial posts across all analyses, showing that many users tend to avoid overt crime disclosure and instead anchor their activity in ambiguous content. Such grey content may function as a proxy for implicit crime disclosure, especially considering that grey-dominant users are disproportionately represented in private messaging flows. Navigating in grey areas could be understood as a risk-avoidance behavior \cite{bada_understanding_2021}: users may intentionally provide just enough information to attract potential customers while remaining vague enough to maintain plausible deniability or avoid drawing law enforcement attention. This is possible given the neutrality of IT, which allows many cybercrime-related tasks to appear legitimate at first glance~\cite{bijlenga_criminals_2018}. The large proportion of grey posts could also be explained by broader trends of commoditization and specialization in the cybercrime ecosystem~\cite{moore_economics_2009, wegberg_plug_2018, collier_cybercrime_2020, collier_cybercrime_2021}. Further research investigating and categorizing grey posts could help clarify the forms of \textit{implicit} criminal activity they may conceal.

    Finally, similar to prior research that used similar techniques to label large quantities of underground forum posts and threads~\cite{clairoux-trepanier_use_2024, moreno-vera_beneath_2025}, this study leverages LLM to label nearly 122,000 initial thread posts. The results highlight a great potential for using LLMs to categorize and/or label cybercrime forum posts into abstract concepts, such as crime disclosure. As stated in the methodology, the coherence and accuracy of the LLM showed good results: average pairwise agreement of 95\% and accuracy in the labelling of 83\%. Because there remains room to further improve LLM performance, one might argue in favour of human annotators for such complex labelling tasks. However, prior research also shows that human labelling is far from flawless, with inconsistencies even among trained coders~\cite{singh_students_2022}. With ongoing advances in LLM research and continued improvements in their application to labelling tasks, LLMs represent a valuable tool for extracting higher-level concepts from cybercrime forum discussions.

    Moreover, as shown in earlier research~\cite{morini_participant_2025, Vermeer_Trilling_2020}, Markov chains are well suited for analysing user (posting) behavior which unfolds through a sequence. This study shows the value of this tool for capturing user behavior patterns in cybercrime forums. While our analysis focused on identifying crime disclosure patterns, future work could extend this approach toward predictive modelling to forecast user posting behaviors within such environments.

    \subsection{Limitations}
        This study has several limitations. We only labelled crime disclosure in initial thread posts, not in comments or private messages. Further research could examine disclosure patterns by incorporating posts beyond thread initiating ones.  This might lead to an increase in the proportion of users disclosing crime. Also, we examined a single forum over 1.5 years, which limits generalizability and may introduce temporal biases. Some posts may have been removed before data collection, potentially skewing results toward persistent users. Studying crime disclosure patterns in several cybercrime forums would overcome this limit and yield different, interesting results. 
        
        Moreover, because the Markov chain analysis relies solely on the present disclosure state, it does not capture longer-term temporal dynamics in user behavior. Subsequent work could focus on modelling user trajectories of crime disclosure over time.
        
        The LLM-based labelling achieved 83\% accuracy on a test subset, with misclassifications typically differing by only one category and showing minimal disagreement across multiple runs. However, it is worth noting that distinguishing among the four labels (unclear, benign, grey, and crime-related) proved challenging in practice, and although we established detailed criteria, these definitions remain imperfect and open to refinement in future work. More generally, it is worth highlighting that our definition of ``crime-related'' activity does not necessarily correspond to criminal behavior in a legal sense, as we lacked ground-truth data linking crime disclosure in a cybercrime forum to actual criminal actions or convictions.

\section{Conclusion}
    In this study, we examine crime disclosure patterns in a large cybercrime forum comprising over 3.5 million posts and nearly 300,000 users. Participants' posts are classified into three disclosure levels: (1) benign, (2) grey, and (3) crime-related, allowing us to capture both explicit and ambiguous forms of criminal expression. Using a combination of LLM-based classification, Markov-chain modelling, logistic regression, and assortative-mixing analysis, this study provides the first systematic assessment of crime disclosure dynamics in a prominent cybercrime forum. Methodologically, the study advances current research by applying LLMs to label large quantities of underground forum discussions.

    Future research should extend this approach to a broader set of cybercrime forums to examine how disclosure patterns vary across them. Such comparative work would help identify which forums are more overtly exposed than others and move beyond the binary distinction between ``cybercrime'' and ``non-cybercrime'' forums by recognizing that crime exposure varies substantially between communities. Moreover, the prominence of grey topics warrants deeper investigation to better understand how and why individuals posting grey content may refrain from overt crime disclosure or, alternatively, may not be involved in crime at all.

\section{Ethical Considerations}
    Given that the study data originated from a forum database disclosed publicly several years prior, rigorous ethical considerations are essential. To ensure responsible data handling and to further uphold ethical standards in using LLMs for this study, we adhered to the principles articulated in the Menlo Report \cite{kenneally_menlo_2012}. Furthermore, ethical approval was obtained from the institutional ethics committees of all affiliated research institutions (University of Twente and Université de Montréal).

    \textbf{Respect for Persons:} The dataset contains information potentially linking online personas to individuals, and was obtained through a data breach without user consent. To mitigate this, we have removed as much directly identifying personal information as possible and avoided showing any personally identifiable details.
    \textbf{Beneficence:} We believe the risk of harm to individuals whose data is in this dataset is minimal. While our research may raise awareness of the dataset's existence, we consider this ethically justifiable given the breach occurred approximately ten years ago and we are not referencing the original source.
    \textbf{Justice:} To minimize selection bias, all threads within the dataset were treated equally throughout our analysis.
    \textbf{Respect for Law and Public Interest:} We are committed to transparency in our research methods and provide clear explanations of our data processing and analysis.

\section*{Author Contribution Statement}
    \textbf{R. Hoheisel}: conceptualization, writing, reviewing, analysis, methodology, investigation. \textbf{T. Meurs}: conceptualization, methodology, investigation, reviewing. \textbf{J. Wientjes:} conceptualization, methodology, investigation, reviewing. \textbf{M. Junger}: conceptualization, reviewing. \textbf{A. Abhishta}: conceptualization, reviewing. \textbf{M. Paquet-Clouston}: conceptualization, writing, reviewing, analysis, methodology, investigation.

\begin{acks}
    We are very thankful to the Dutch Police for their support. The opinions expressed in this work are ours and do not necessarily represent the views of the Dutch Police. We also thank the reviewers for their valuable feedback. The labelling of data for this research was done using OpenAI GPT-4.1. This manuscript’s English was partially revised using ChatGPT (OpenAI) for grammar and style. 
\end{acks}

\bibliographystyle{ACM-Reference-Format}
\bibliography{drift_crime,refs}

\appendix

\section{Labelling Prompts}
     To optimize prompts, we tested various strategies to maximize accuracy against ground truth data without being too specific. Our approach was informed by existing literature and relevant publications \cite{roy_darkgram_2025, moreno-vera_beneath_2025, clairoux-trepanier_use_2024}. We experimented with information density (from minimal guidelines to detailed instructions with examples), chain-of-thought prompts that decomposed labeling into sequential reasoning steps, and instruction placement across system versus user prompts to assess performance impacts. Figure \ref{app:system_prompt} shows the system prompt and Figure \ref{app:user_prompt} the user prompt used for the disclosure labelling.
\newpage
\begin{figure*}[htb]
        \footnotesize
        \begin{lstlisting}
        You are a content classification assistant trained to analyze posts from underground forums.
        
        Your role is to evaluate whether a post is connected to criminal activity.
        
        Important rules:
        - Ignore third-party content (e.g., reposted news articles or opinions about external crimes).
        - When in doubt, choose the less severe category, but only if there is no direct indication of 
        criminal intent or context.
        - Do not speculate beyond what is stated or clearly implied.
        \end{lstlisting}
        \caption{System prompt used for the crime disclosure labelling.}
        \label{app:system_prompt}
    \end{figure*}
    
    \begin{figure*}[htb]
        \footnotesize
        \begin{lstlisting}
        You are classifying a forum post based on its connection to criminal activity based on its content, title, 
        the summarized comments, and the sub-forum it was posted in.
        
        Follow these steps carefully before deciding:
        1. Summarize the core purpose or action of the post. Take also the comment summary, the sub-forum and the title of
        the post into account.
        2. Determine whether the post promotes, facilitates, or discusses criminal behavior. 
        3. Evaluate whether the behavior is clearly illegal, context-dependent (grey), or benign. The comment summary and 
        sub-forum name may provide additional information. Use the following label descriptions:
        - Crime related: Assign this label if the post: 
            - Mentions cracked or pirated software, mods, or exploits - including download links, sales, or requests.
            - Promotes fraudulent money-making schemes, such as suspicious "get rich quick" games, referral scams, 
            or giveaways requiring registration for unknown rewards
            - Offers or discusses hacking tools, malware, phishing, or other clearly illegal services.
        - Grey: Assign this label if the post:
            - Assign this label if the post: Shares, buys, sells, or requests accounts, or credentials, even if the 
            service itself is not cracked or illegal. 
            - Discusses cheats, scripts, assemblies, or botting tools for games
            - Promotes discount services or gift card shops that feel suspicious but don't show clear signs of fraud.
            - Offers giveaways that aren't obvious scams.
        - Benign: Assign this label if the post:
            - Provides legitimate how-to guides, modding tutorials, or game tips.
            - Asks about or explains game features, forum reputation, general tech/gaming topics.
            - Mentions official software or promotions without involving cracks, malware, cheats, or credentials.
            - Examples: General tech help, gameplay questions, opinions;Personal content, non-suspicious buying/selling of 
            own services;Forum mechanics (e.g., access errors, reputation systems)
        - Unclear: Assign this label if the post:
            - There's not enough context to confidently apply another label.
        4. Conclude with one of: 'crime related', 'grey', 'benign', or 'unclear'.        
    
        ---
        
        Now classify this post:
        
        Sub-forum: {}
        Post Title: {}
        Post Content: {}
        Comment Summary: {}
    
        ---
        
        Respond with ONLY the final label.
        \end{lstlisting}
        \caption{User prompt used for the crime disclosure labelling.}
        \label{app:user_prompt}
    \end{figure*}

\end{document}